\documentclass[12pt,twocolumn,reqno]{article}
\usepackage{balance}
\usepackage{amsmath}
\usepackage{graphicx}
\usepackage{color}
\begin{document}

\twocolumn[
\begin{@twocolumnfalse}
\title
\centerline{\bf\large A New Interpretation of Charge Based on Spinning Sphere Model of Electron}

\vspace{12pt}

\centerline{{\bf S. Ghosh}$^{\rm 1}$, {\bf A. Choudhury}$^{\rm 2}$ and  {\bf J. K. Sarma }}
\centerline{Department of Physics, Tezpur University, Tezpur, Assam, 784 028, India} 
\centerline{{{\it gsovan@gmail.com}$^{\rm 1}$}, {{\it ajc@tezu.ernet.in}$^{\rm 2}$}}

\begin{abstract}
\leftskip1.0cm
\rightskip1.0cm

LEP experiments indicate that the charge of the electron is distributed over a small radius $\sim10^{-20} $m. By incorporating this information in spinning sphere model of electron we arrive at a new interpretation of charge as the Schwinger corrected or electromagnetic or magnetic self-energy mass. The charge-energy equivalence is also arrived at in this paper. Further it is predicted that particles with zero mass will not contain any charge.

\end{abstract}
\end{@twocolumnfalse}
]

\paragraph \ 
LEP experiments indicate that the charge of the electron is distributed over a small radius $\sim 10^{-20}$ m [1]. The explanation of related scattering by QED[2,3] too demand that the charge of electron is concentrated with a smaller mass as compared to total mass of electron. This prompts us to probe any possible link between mass and charge of an elementary particle. The similarity of inverse square law of gravitation and charge also forces us to explore this link.

\ The experimental result of magnetic moment of electron is not found to match with the magnetic moment when only the mechanical mass of the electron is considered[4,5]. Schwinger gave a correction term $m.\frac{\alpha}{2\pi}$, where $\alpha = \frac{e^2}{\hbar c}$[4,5](dimensionless constant) to compensate for the difference between the two. This correction term is known as the electromagnetic mass of electron[4,5]. It may be noted that $\alpha$ is the so called fine structure constant coupling the strength of interaction between electron and photon[6]. 

\ In the standard SS (Spinning Sphere) model [5] of the electron, let us assume that the charge is confined in a very small region of the electron. This assumption, it may be noted that, is consistent with the demands of LEP experiments and QED theory. Thus we allow the charge-part of electron to be a small sphere of radius $R_E$ which is smaller than the corresponding classical radius, $R_0$ of the electron.

\ The mass density of the sphere of radius $R_0$ and total mass $m(1+\frac{\alpha}{2\pi})$ is 

\begin{equation}D_0  = \frac{m.\frac{\alpha}{2\pi}(1+\frac{2\pi}{\alpha})}{\frac{4}{3} \pi {R_0}^3}\end{equation} 

\ The ratio of the charge densities of the bigger sphere to that of the sphere containing electromagnetic mass is

\begin{equation}\frac{{\rho}_0}{{\rho}_E} = \frac{{R_E}^3}{R_0^3}\end{equation} 

From (1) and (2) we have 

\begin{displaymath}D_0 = \frac{m.\frac{\alpha}{2\pi}}{\frac{4}{3} \pi R_E^3} (1+\frac{2\pi}{\alpha}) \frac{\rho_0}{\rho_E}\end{displaymath}

Therefore

\begin{equation}\frac{D_0}{D_E} = (1+\frac{2\pi}{\alpha})\frac{\rho_0}{\rho_E}\end{equation}
where $D_E$ is the density of the smaller sphere containing electromagnetic mass.

\ Electromagnetic mass of the electron divided by the volume of the smaller charge-sphere in equation (3) provides us the mass density ($D_E$) of the charge part. Hence the ratio of the mass densities is proportional to the ratio of the charge densities

\begin{equation} \frac{D_0}{D_E} = Const. (1+\frac{2\pi}{\alpha}) \frac{\rho}{\rho_1}\end{equation}

\ Now if we set \begin{equation} D = (1+\frac{2\pi}{\alpha})\rho \end{equation}
as the $\frac{\alpha}{2\pi}$ term may dominate over other constant parameters, we have

\begin{equation}\frac{mass}{volume} = (1+\frac{2\pi}{\alpha})\frac{charge}{volume}\end{equation}
Again as $\frac{2\pi}{\alpha}>>1$ approximating equation (6) we have 

\begin{equation} e = m. \frac{\alpha}{2\pi}\end{equation}

\ Thus the charge of the electron in this model can be identified as the Schwinger corrected or electromagnetic or self-energy mass. 

\ Further magnetic self-energy of the electron is given by [5][7]

\begin{equation} W_H = m.\frac{\alpha}{2\pi} c^2\end{equation}

\ With reference to (7), \begin{equation} W_H = ec^2\end{equation} 

\ It is the charge-energy equivalence relation which puts charge on the same footing as mass (as in $E=mc^2$). 

\ Equation (7) implies that the measure of charge for zero-mass particle is zero itself. This finding is in agreement with the experimental observations as of today.

\ We note that the total energy of the electron consists of the contribution from mass and charge component;i.e.  
\begin{equation}E = mc^2 +m.\frac{\alpha}{2\pi} c^2 = (m+e)c^2\end{equation}

{\bf{References}}

\ 1. M. Rivas,J. Phys. A: Math. Gen 36 4703-4715(2003) 

\ 2. F. Halzen and A. D. Martin, Quarks and leptons, John Wiley and Sons (1984)

\ 3. D. Griffiths, Introduction to elementary particles, John Wiley and Sons (1987)

\ 4. J. Schwinger, Phys. Rev. 73, 416 (1948)

\ 5. M. H. MacGregor, The Enigmatic Electron, Kluwer Academic Publishers, Dordrecht (1992).

\ 6. V. F.Weiskopf, Phys. Rev., 56, 72 (1939)

\ 7. M. H. MacGregor, Found. Phys. Lett., 2, 577 (1989)

\end{document}